\documentclass[runningheads,a4paper]{llncs}
\usepackage{amssymb}
\usepackage{graphicx}
\usepackage{listings}
\usepackage{url}
\usepackage{cite}
\urldef{\mailsa}\path|{hany, mln}@cs.odu.edu|    
\newcommand{\keywords}[1]{\par\addvspace\baselineskip
\noindent\keywordname\enspace\ignorespaces#1}
\usepackage{colortbl}
\usepackage{alltt}
\begin{document}

\mainmatter

\title{Losing My Revolution}
\subtitle{How Many Resources Shared on Social Media\\ Have Been Lost?}

\author{Hany M. SalahEldeen\and Michael L. Nelson}

\institute{Old Dominion University, Department of Computer Science\\
Norfolk VA, 23529, USA\\
\mailsa\\}

%
%

\maketitle

\begin{abstract}
Social media content has grown exponentially in the recent years and the role of social media has evolved from just narrating life events to actually shaping them. In this paper we explore how many resources shared in social media are still available 
on the 
live web or in public web archives. By analyzing six different event-centric datasets of resources shared 
in social media in the period from June 2009 to March 2012, we found about 11\% lost and 20\% archived after just a year and an average of 27\% lost and 41\% archived after two and a half years. Furthermore, we found a nearly 
linear relationship between time of sharing of the 
resource and the percentage lost, with a slightly less linear relationship between time of sharing and archiving coverage of the resource. From this model we conclude that after the first year of publishing, nearly 11\% of shared resources 
will be lost and after that we will continue to lose 0.02\% per day.

\end{abstract}

\section{Introduction}
With more than 845 million Facebook users at the end of 2011 \cite{facebook} and over 140 million tweets 
sent daily in 2011 \cite{Twitternums} users can take photos, videos, post their opinions, and report incidents as they happen. 
Many of the posts and tweets are about quotidian events and their preservation is debatable. However, some of the posts and events are about culturally important events whose preservation is less controversial. In this paper we shed light on the importance of 
archiving social media content about these events and estimate how much of this content is archived, still available, or lost with no 
possibility of recovery.\\

To emphasize the culturally important commentary and sharing, we collected data about six events in the time period of June 2009 to March 2012: the H1N1 virus outbreak, Michael Jackson's death, the Iranian elections and protests, 
Barack Obama's Nobel Peace Prize, the Egyptian revolution, and the Syrian uprising.

\section{Related Work}
To our knowledge, no prior study has analyzed the amount of shared resources in social media lost through time. There have been many studies analyzing the behavior of users within a social network, how they interact, and what content they 
share \cite{Benevenut,Wu,Zhao,Wilson}. As for Twitter, Kwak et al. \cite{Kwak} studied its nature and its topological characteristics and found a deviation from known characteristics of human social networks that were analyzed by Newman and Park \cite{human}. Lee analyzed 
the reasons behind sharing news in social media and found that informativeness was the 
strongest motivation in predicting news sharing intention, followed by socializing and status seeking \cite{Chei}. Also shared content in social media like Twitter move and diffuse relatively fast as stated by Yang et al. \cite{Yang}.\\

 Further more, many concerns were raised about the persistence of shared resources and web content in general. Nelson and Allen studied the persistence of objects in a digital library and found that, with just 
over a year, 3\% of the sample they collected have appeared to no longer be available \cite{nelson1}. Sanderson et al. analyzed the persistence and availability 
of web resources referenced from papers in scholarly repositories using Memento and found that 28\% of these resources have been lost \cite{Sanderson}. Memento \cite{memento} is a collection of HTTP extensions that enables uniform, inter-archive access. Ainsworth et al. \cite{Ainsworth:2011:MWA:1998076.1998100} examined how 
much of the web is archived and found it ranges from 16\% to 79\%, 
depending on the starting seed URIs.  McCown et al. examined the factors 
affecting reconstructing websites (using caches and archives) and found 
that PageRank, Age, and the number of hops from the top-level of the
site were most influential \cite{1255182}.

\section{Data Gathering}
We compiled a list of URIs that were shared in social media and correspond to specific culturally important events. In this section we describe the data acquisition and 
sampling process we performed to extract six different 
datasets which will be tested and analyzed in the following sections.

\subsection{Stanford SNAP Project Dataset}
The Stanford Large Network Dataset is a collection of about 50 large network datasets having millions of nodes, edges and tuples. It was collected as a part of the Stanford Network Analysis Platform (SNAP) project \cite{snap}. It includes 
social networks, web graphs, road networks, Internet networks, citation networks, collaboration networks, and communication networks. For the purpose of our investigation, we selected their Twitter posts dataset. 
This dataset was collected from June 1st, 2009 to December 31st, 2009 and contains nearly 476 million tweets posted by nearly 17 million users. The dataset is estimated to cover 20\%-30\% of all posts published on Twitter during that time 
frame\cite{Twitter}. To select which events will be covered in this study, we examined CNN's 2009 events timeline\footnote{http://www.cnn.com/2009/US/12/16/year.timeline/index.html}. We wanted to select a small number of events that were diverse, with limited overlap, and relatively important to a large number of people. Given that, we selected four events: the H1N1 virus outbreak, the Iranian protests and 
elections, Michael Jackson's death, and Barrack Obama's Nobel Peace Prize award.

\subsubsection{Preparation:} A tweet is typically composed of text, hashtags, embedded resources or URIs and usertags all spanning a maximum of 140 characters. Here is an example of a tweet record in the SNAP dataset: 

\begin{minipage}[b]{\textwidth}
\scriptsize
\vskip 5pt
\textbf{T}	2009-07-31 23:57:18\\
\textbf{U}	http://Twitter.com/nickgotch\\
\textbf{W}	RT @rockingjude: December 21, 2009 Depopulation by Food Will Begin \\http://is.gd/1WMZb WHOA..BETTER WATCH RT plz \#pwa \#tcot\\ \vskip -5pt
\normalsize
\end{minipage}

The line starting with the letter \textbf{T} indicates the date and time of the tweet creation. While the line starting with \textbf{U} shows a link to the user who authored this particular tweet. Finally, the line starting with \textbf{W} shows the entire 
tweet including all the user-references ``@rockingjude'', the embedded URIs ``http://is.gd/1WMZb'', and hashtags ``\#pwa \#tcot''.
\subsubsection{Tag Expansion:} We wanted to select tweets that we can say with high confidence are about a selected event. In this case, precision is more important than recall as collecting every single tweet published about a certain event is less important than making sure that the selected tweets are definitely about that event. Several studies 
focused on estimating the aboutness of a certain web page or a resource in general \cite{lexical,lexical2}. Fortunately in Twitter, hashtags incorporated within a tweet can help us estimate their ``\textit{aboutness}''. Users normally add certain hashtags to their tweets to ease 
the search and discoverability in following a certain topic. These hashtags will be utilized in the event-centric filtration process.\\

For each event, we selected initial tags that describe it (Table \ref{tab:tags}). Those initial tags were derived empirically after examining some event-related tweets. 
Next we extracted all the hashtags that co-occurred with our initial set of hashtags. For example, in class H1N1 we extracted all the other hashtags that appeared along with \textit{\#h1n1} within the same tweet 
and kept count of their frequency. Those extracted hashtags were sorted in descending order of the frequency of their appearance in tweets.  We removed all the general scope tags like \textit{\#cnn}, \textit{\#health}, 
\textit{\#death}, \textit{\#war} and others. In regards to aboutness, removing general tags will indeed decrease recall but will increase precision. Finally we picked the top 8-10 hashtags to represent this event-class and 
be utilized in the filtration process. Table \ref{tab:tags} shows the final set of tags selected for each class.
\begin{center}
\vskip-21pt
  \begin{table}
      \begin{tabular}[c]{ | p{1.5cm} || p{2.2cm} | p{8.1cm} |}
      \hline  
      \scriptsize\textbf{Event} & \scriptsize \textbf{Initial Hashtags} &\scriptsize \textbf{Top Co-occurring Hashtags} \\ \hline
      \tiny \textbf{H1N1} &\tiny `h1n1' &\tiny `swine'=61,829 `swineflu'=56,419 `flu'=8,436\\
      \tiny \textbf{Outbreak} &\tiny =61,351 & \tiny  `pandemic'=6,839 `influenza'=1,725 `grippe'=1,559 `tamiflu'=331 \\ \hline
      \tiny \textbf{M. Jackson's}&\tiny  `michaeljackson' &\tiny `michael'=27,075 `mj'=18,584 `thisisit'8,770 `rip'=3,559 `jacko'=3,325  \\
      \tiny \textbf{Death} &\tiny=22,934  &\tiny`kingofpop'=2,888 `jackson'=2,559  `thriller'=1,357 `thankyoumichael'=1,050\\ \hline
      \tiny \textbf{Iranian}&\tiny  `iranelection' &\tiny `iran'949,641 `gr88'=197,113`tehran'=109,006 `freeiran'=13,378\\ 
      \tiny \textbf{Elections} &\tiny  =911,808 &\tiny `neda'=191,067  `mousavi'=16,587  `united4iran'=9,198 `iranrevolution'=7,295\\ \hline
      \tiny \textbf{Obama's} &\tiny  `obama'=48,161 \& &\tiny `nobel'=2,261 `obamanobel'=14 `nobelprize' `nobelpeace'=113 \\
      \tiny \textbf{Nobel Prize} &\tiny  `peace`=3,721 &\tiny `barack'=1292 `nobelpeaceprize'=107\\ \hline
      \end{tabular}\\ \normalsize 
      \caption{Twitter hashtags generated for filtering and their frequency of occurring}
      \label{tab:tags}
\vskip-25pt
  \end{table}
\end{center}
\subsubsection{Tweet Filtration:} In the previous step we extracted the tags that will help us classify and filter tweets in the dataset according to each event. This filtration process aims to extract a reasonable sized dataset of tweets 
for each event and to minimize the inter-event overlap. Since the life and persistence of the tweet itself is not the focus of this study but rather the associated resource that appears in the tweet (image, video, shortened URI or other 
embedded resource), we will extract only the tweets that contain an embedded resource. This step resulted in 181 million tweets with embedded resources (http://is.gd/1WMZb in the prior example). These tweets were further filtered to keep only the tweets that have at least one of 
the expanded tags obtained from Table \ref{tab:tags}. The number of tweets after this phase reached 1.1 million tweets.\\

Filtering the tweets based on the occurrence of at least one of the hashtags only is undesirable as it will cause two problems: First, it will introduce possible event overlap due to general tweets talking about two or more topics. Second, is that 
using only the single occurrence of these tags will yield a huge amount of tweets and we need to reduce this size to reach a more manageable size.  Intuitively speaking, strongly related hashtags will co-occur often. For example, a tweet that has \textit{\#h1n1} along with \textit{\#swineflu} and \textit{\#pandemic} is most likely about the H1N1 outbreak rather than a tweet having just the tag \textit{\#flu} or just \textit{\#sick}. 
Filtering with this co-occurrence will in turn solve both problems as by increasing relevance to a particular event, general tweets that talk about several events will be filtered out thus diminishing the overlap, and in turn it will reduce the size of the dataset.\\

Next, we increase the precision of the tweets associated with each event from the set of 1.1 million tweets. In the first iteration we selected the tag that had the highest frequency of co-occurrence in the dataset with the initial tag and added it to a set we will call the selection set. After that we check the co-occurrence of all the remaining 
extracted tags with the tag in the selection set and record the frequencies of co-occurrence. After sorting the frequencies of co-occurrence with the tag from the selection set, we pick the highest one to keep add it to the selection set. We repeat this 
step of counting co-occurrences but with all the previously extracted hashtags in the selection set from previous iterations.\\

To elaborate, for H1N1 assume that the hastag `\#h1n1' had the highest frequency of appearance in the dataset so we add 
it to the selection set. In the next iteration we record the how many times each tag in the list appeared along with `\#h1n1' in a same tweet. If we selected `\#swine' as the one with the highest frequency of occurrence with the initial tag `\#h1n1' we add 
it to the selection list and in the next iteration we record the frequency of occurrence of the remaining hashtags with both of the extracted tags `\#h1n1' and `\#swine'. We repeat this step, for each event, to the point where we have a manageable size 
dataset which we are confident in its `aboutness' in relation to the event.
\renewcommand{\arraystretch}{0.6}
\begin{center}
  \begin{table}
      \vskip-20pt
      \begin{tabular}[c]{ | p{0.8cm} || p{4.7cm} | p{2.1cm} | p{2.5cm} | p{1.5cm} |}
      \hline  
      \tiny  \textbf{Event} & \tiny \textbf{Hashtags selected for filteration} & \tiny \textbf{Tweets Extracted} &  \tiny \textbf{Operation Performed} & \tiny \textbf{Final Tweets}\\ \hline \hline
      \tiny  \textbf{MJ} & \tiny michael & \tiny 27,075 & &\\ 
      \tiny  & \tiny michael \& michaeljackson & \tiny \textbf{22,934} & \tiny Sample 10\% &\tiny \textbf{2,293} \\ \hline
      \tiny  \textbf{Iran} & \tiny iran & \tiny 949,641  & & \\ 
      \tiny   & \tiny iran \& iranelection & \tiny 911,808 &  & \\ 
      \tiny   & \tiny iran \& iranelection \& gr88  & \tiny 189,757 &  & \\ 
      \tiny   & \tiny iran \& iranelection \& gr88 \& neda  & \tiny 91,815 &  & \\ 
      \tiny   & \tiny iran \& iranelection \& gr88 \& neda \& tehran  & \tiny \textbf{34,294} &  \tiny Sample 10\%  &\tiny \textbf{3,429} \\ \hline
      \tiny  \textbf{H1N1} & \tiny h1n1 & \tiny 61,351 &  &\\ 
      \tiny  & \tiny h1n1 \& swine & \tiny 44,972  &  &\\ 
      \tiny  & \tiny h1n1 \& swine \& swineflu  & \tiny 42,574 &  &\\ 
      \tiny  & \tiny h1n1 \& swine \& swineflu \& pandemic  & \tiny \textbf{5,517}  &  \tiny Take All& \tiny \textbf{5,517}\\ \hline
      \tiny  \textbf{Obama} & \tiny obama & \tiny 48,161 &  &\\ 
      \tiny  & \tiny obama \& nobel & \tiny \textbf{1,118} & \tiny Take All & \tiny \textbf{1,118} \\ \hline
      \end{tabular}\\ \normalsize 
      \caption{Tweet Filtration iterations and final tweet collections}
      \vskip-45pt
      \label{tab:tagoccurrence}
  \end{table}
\end{center}
\renewcommand{\arraystretch}{1}

Two problems appeared from this approach with the Iran and Michael Jackson datasets. In the Iran dataset the number of tweets was in hundreds of thousands and even with 5 tags co-occurrence it was still about 34K+ tweets. To solve this we performed a random 
sampling from those resulting tweets to take only 10\% of them resulting in a smaller manageable dataset. The second problem with the Michael Jackson dataset upon using 5 tags to decrease it to a manageable size we realized there were 
few unique domains for the embedded resources. A closer look revealed this combination of tags was mostly border-line tweet spam (MJ ringtones). To solve this we used only the two top tags ``\#michael'' and ``\#michaeljackson'', and then we randomly sampled 10\% of the resulting tweets to reach 
the desired dataset size (Table \ref{tab:tagoccurrence}).

\subsection{Egyptian Revolution Dataset}
The one year anniversary of this event was the original motivation for this study \cite{losingrevolution}. In this case, we started with an event and then tried to get social media content describing it. Despite its ubiquity, gathering social media for a past 
event is surprisingly hard. We picked the Egyptian revolution due to the role of the social media in curating and driving the incidents that led to the resignation of the president. 
Several initiatives were commenced to collect and curate the social media content during the revolution like R-sheif.org\footnote{http://www.r-shief.org/} which specializes in social content analysis of the issues in the Arab world by using 
aggregate data from Twitter and the Web. We are currently in the process of obtaining the millions of records related to the Arab Spring of 2011. Meanwhile, we decided to build our own dataset manually.\\

There are several sites that curate resources about the Egyptian Revolution and we want to investigate as 
many of them as possible. At the same time, we need to diversify our resources and the types of digital artifacts that are embedded in them. Tweets, videos, images, embedded links, entire web pages and books were included in our investigation. 
For the sake of consistency, we limited our analysis to resources created within the period from the 20th of January 2011 to the 1st of March 2011. In the next subsections 
we explain each of the resources we utilized in our data acquisition in detail.
\subsubsection{Storify:}
Storify is a website that enables users to create stories by creating collections of URIs (e.g., Tweets, images, videos, links) and arrange them temporally. These entries are posted by reference to their 
host websites. Thus, adding content to Storify does not necessarily mean it is archived. If a user added a video from YouTube and after a while the publisher of that video decided to remove it from YouTube 
the user is left with a gap in their Storify entry. For this purpose we gathered all the Storify entries that were created between 20th of January 2011 and the 1st of March 2011, resulting in 219 unique resources.

\subsubsection{IAmJan25:}
Some entire websites were dedicated as a collection hub of media to curate the revolution. Based on public contributions, those websites collect different types of media, classify them, order them chronologically and publish them to the public. 
We picked a website named IAmJan25.com, as an example of these websites, to analyze and investigate. 
The administrators of the website received selected videos and images for notable events and actions that happened during the revolution. 
Those images and videos were selected by users as they vouched for them to be of some importance and they send the resource's URI to the web site administrators. The website itself is divided into 
two collections: a video collection and an image collection. The video collection had 2387 unique URIs while the image collection had 3525 unique URIs.

\subsubsection{Tweets From Tahrir:}
Several books were published in 2011 documenting the revolution and the Arab Spring. To bridge the gap between books and digital media we analyzed a book entitled \textbf{\textit{Tweets from Tahrir}} \cite{tweetsfromtahrir} which was published on April 21st, 2011. 
As the name states, this book tells a story formed by tweets of people during the revolution and the clashes with the past regime. We analyzed this book as a collection of tweets that had the luxury of a paperback preservation and focused on the 
tweeted media, in this case images. The book had a total of 1118 tweets having 23 unique images.

\subsection{Syria Dataset}
This dataset has been selected to represent a current (March 2012) event. Using the Twitter search API, we followed the same pattern of data acquisition as in section 3.1. We started with one 
hashtag, \#Syria, and expanded it. Table \ref{tab:syriatags} show the tags produced from the tag expansion step. After that each of those tags were input into a process utilizing the Twitter streaming API and produced the first 1000 
results matching each tag. From this set, we randomly sampled 10\%. As a result, 1955 tweets were extracted each having one or more embedded resources and tags from the expanded tags in Table \ref{tab:syriatags}.
\begin{center}
  \begin{table}
\vskip-28pt
      \begin{tabular}[c]{ |p{2.2cm} || p{9.8cm} |}
      \hline  
      \scriptsize \textbf{Initial Hashtags} &\scriptsize \textbf{Extracted Hashtags} \\ \hline
      \tiny `Syria' &\tiny `Bashar' `RiseDamascus' `GenocideInSyria' `STOPASSAD2012' `AssadCrimes' `Assad'\\ \hline
      \end{tabular}\\ \normalsize 
      \caption{Twitter \#Tags generated for filtering the Syrian uprising}
      \label{tab:syriatags}
  \end{table}
\vskip-50pt
\end{center}

Table \ref{tab:domains} shows the resources collected along with the top level domains that those resources belong to for each event.
\begin{center}
  \begin{table}
\vskip-30pt
      \begin{tabular}[c]{ |p{1.3cm}|| p{10.7cm}|}
      \hline  
      \scriptsize \textbf{ Event }  & \scriptsize \textbf{ Top Domains (number of resources found) }  \\ \hline \hline
      \scriptsize MJ  & \scriptsize youtube (110), twitpic (45), latimes (43), cnn (30), amazon (30)\\ \hline 
      \scriptsize Iran  & \scriptsize youtube (385), twitpic (36), blogspot (30), roozonline (29)\\ \hline 
      \scriptsize H1N1  & \scriptsize rhizalabs (676), reuters (17), google (16), flutrackers (16), calgaryherald (11)\\ \hline 
      \scriptsize Obama  & \scriptsize blogspot (16), nytimes (15), wordpress (12), youtube (11), cnn (10)\\ \hline 
      \scriptsize Egypt  & \scriptsize youtube (2414), cloudfront (2303), yfrog (1255), twitpic (114), imageshack.us (20)\\ \hline 
      \scriptsize Syria  & \scriptsize youtube (130), twitter (61), hostpic.biz (9), telegraph.co.uk (5)\\ \hline 
      \end{tabular}\\ \normalsize 
      \caption{The top level domains found for each event ordered descendingly by the number of resources.}
      \label{tab:domains}
\vskip-30pt
  \end{table}
\vskip-25pt
\end{center}

\section{Uniqueness and Existence}
From the previous data gathering step we obtained six different datasets related to six different historic events. For each event we extracted a list of URIs that were shared in tweets or uploaded to sites like Storify or IAmJan25. To answer the question of how much of the social media content is missing we test those URIs for each dataset to eliminate URI aliases in 
which several URIs identify to the same resource. Upon obtaining those unique URIs we examine how many of which are still available on the live web and how many are available in public web archives. 

\subsection{Uniqueness}
Some URIs, especially those that appear in Twitter, may be aliases for the same resource. For example ``http://bit.ly/2EEjBl'' and ``http://goo.gl/2ViC'' both resolve to ``http://www.cnn.com''. To solve this, we resolved all 
the URIs following redirects to the final URI. The HTTP response of the last redirect has a field called \textit{location} that contains the original long URI of the resource. This step reduced the total number of URIs 
in the six datasets from 21,625 to 11,051. Table \ref{tab:all} shows the number of unique resources in every dataset.

\begin{table}[ht]
      \vskip-15pt
\begin{minipage}[b]{0cm}
\centering
\begin{tabular}{r|c|c c}
\multicolumn{1}{r}{}
 &  \multicolumn{1}{c}{\scriptsize All} & \multicolumn{1}{c}{\scriptsize Unique}  & \multicolumn{1}{c}{\scriptsize }\\
\cline{2-3}
\scriptsize  & \tiny 2,293 & \multicolumn{1}{c|}{\cellcolor[gray]{0.8} \tiny 1,187=\textbf{51.77\%} } & \\
\cline{2-3}
\multicolumn{1}{r}{}
\underline{\textbf{\emph{MJ}}} &  \multicolumn{1}{c}{\tiny Archived} & \multicolumn{1}{c}{\tiny Not Archived} & \multicolumn{1}{c}{\scriptsize }\\
\cline{2-3}
\tiny Available & \tiny 316=26.62\% & \multicolumn{1}{c|}{\tiny 474=39.93\% }  & \tiny\\
\cline{2-4}
\tiny Missing & \tiny 90=7.58\% & \multicolumn{1}{c|}{\tiny 307=25.86\% }  & \multicolumn{1}{c|}{\cellcolor[gray]{0.8} \tiny 397=\textbf{33.45\%} }  \\ 
\cline{2-4}
\scriptsize  &\cellcolor[gray]{0.8} \tiny 406=\textbf{34.20\%} & \tiny each/1,187  & \tiny \\
\cline{2-2}
\end{tabular}
\end{minipage}
\hspace{6cm}
\begin{minipage}[b]{0cm}
\centering
\begin{tabular}{r|c|c c}
\multicolumn{1}{r}{}
 &  \multicolumn{1}{c}{\scriptsize All} & \multicolumn{1}{c}{\scriptsize Unique}  & \multicolumn{1}{c}{\scriptsize }\\
\cline{2-3}
\scriptsize  & \tiny 3,429 & \multicolumn{1}{c|}{\cellcolor[gray]{0.8} \tiny 1,340=\textbf{39.08\%} } & \\
\cline{2-3}
\multicolumn{1}{r}{}
\underline{\textbf{\emph{Iran}}} &  \multicolumn{1}{c}{\tiny Archived} & \multicolumn{1}{c}{\tiny Not Archived} & \multicolumn{1}{c}{\scriptsize }\\
\cline{2-3}
\tiny Available & \tiny 415=30.97\% & \multicolumn{1}{c|}{\tiny 586=43.73\% }  & \tiny\\
\cline{2-4}
\tiny Missing & \tiny 101=7.54\% & \multicolumn{1}{c|}{\tiny 238=17.76\% }  & \multicolumn{1}{c|}{\cellcolor[gray]{0.8} \tiny 339=\textbf{25.30\%} }  \\ 
\cline{2-4}
\scriptsize  &\cellcolor[gray]{0.8} \tiny 516=\textbf{38.51\%} & \tiny  each/1,340  & \tiny \\
\cline{2-2}
\end{tabular}
\end{minipage}\\
\begin{minipage}[b]{0cm}
\centering
\begin{tabular}{r|c|c c}
\multicolumn{1}{r}{}
 &  \multicolumn{1}{c}{\scriptsize All} & \multicolumn{1}{c}{\scriptsize Unique}  & \multicolumn{1}{c}{\scriptsize }\\
\cline{2-3}
\scriptsize  & \tiny 5,517 & \multicolumn{1}{c|}{\cellcolor[gray]{0.8} \tiny 1,645=\textbf{29.82\%} } & \\
\cline{2-3}
\multicolumn{1}{r}{}
\underline{\textbf{\emph{H1N1}}} &  \multicolumn{1}{c}{\tiny Archived} & \multicolumn{1}{c}{\tiny Not Archived} & \multicolumn{1}{c}{\scriptsize }\\
\cline{2-3}
\tiny Available & \tiny 595=36.17\% & \multicolumn{1}{c|}{\tiny 656=39.88\% }  & \tiny\\
\cline{2-4}
\tiny Missing & \tiny 98=5.96\% & \multicolumn{1}{c|}{\tiny 296=17.99\% }  & \multicolumn{1}{c|}{\cellcolor[gray]{0.8} \tiny 394=\textbf{23.95\%} }  \\ 
\cline{2-4}
\scriptsize  &\cellcolor[gray]{0.8} \tiny 693=\textbf{42.12\%} & \tiny each/1,645  & \tiny \\
\cline{2-2}
\end{tabular}
\end{minipage}
\hspace{6cm}
\begin{minipage}[b]{0cm}
\centering
\begin{tabular}{r|c|c c}
\multicolumn{1}{r}{}
 &  \multicolumn{1}{c}{\scriptsize All} & \multicolumn{1}{c}{\scriptsize Unique}  & \multicolumn{1}{c}{\scriptsize }\\
\cline{2-3}
\scriptsize  & \tiny 1,118 & \multicolumn{1}{c|}{\cellcolor[gray]{0.8} \tiny 370=\textbf{33.09\%} } & \\
\cline{2-3}
\multicolumn{1}{r}{}
\underline{\textbf{\emph{Obama}}} &  \multicolumn{1}{c}{\tiny Archived} & \multicolumn{1}{c}{\tiny Not Archived} & \multicolumn{1}{c}{\scriptsize }\\
\cline{2-3}
\tiny Available & \tiny 143=38.65\% & \multicolumn{1}{c|}{\tiny 135=36.49\% }  & \tiny\\
\cline{2-4}
\tiny Missing & \tiny 33=8.92\% & \multicolumn{1}{c|}{\tiny 59=15.95\% }  & \multicolumn{1}{c|}{\cellcolor[gray]{0.8} \tiny 92=\textbf{24.86\%} }  \\ 
\cline{2-4}
\scriptsize  &\cellcolor[gray]{0.8} \tiny 176=\textbf{47.57\%} & \tiny each/370  & \tiny \\
\cline{2-2}
\end{tabular}
\end{minipage}\\
\begin{minipage}[b]{0cm}
\centering
\begin{tabular}{r|c|c c}
\multicolumn{1}{r}{}
 &  \multicolumn{1}{c}{\scriptsize All} & \multicolumn{1}{c}{\scriptsize Unique}  & \multicolumn{1}{c}{\scriptsize }\\
\cline{2-3}
\scriptsize  & \tiny 7,313 & \multicolumn{1}{c|}{\cellcolor[gray]{0.8} \tiny 6,154=\textbf{84.15\%} } & \\
\cline{2-3}
\multicolumn{1}{r}{}
\underline{\textbf{\emph{Egypt}}} &  \multicolumn{1}{c}{\tiny Archived} & \multicolumn{1}{c}{\tiny Not Archived} & \multicolumn{1}{c}{\scriptsize }\\
\cline{2-3}
\tiny Available & \tiny 1,069=17.37\% & \multicolumn{1}{c|}{\tiny 4440=72.15\% }  & \tiny\\
\cline{2-4}
\tiny Missing & \tiny 173=2.81\% & \multicolumn{1}{c|}{\tiny 472=7.67\% }  & \multicolumn{1}{c|}{\cellcolor[gray]{0.8} \tiny 645=\textbf{10.48\%} }  \\ 
\cline{2-4}
\scriptsize  &\cellcolor[gray]{0.8} \tiny 1242=\textbf{20.18\%} & \tiny each/6,154 & \tiny \\
\cline{2-2}
\end{tabular}
\end{minipage}
\hspace{6cm}
\begin{minipage}[b]{0cm}
\centering
\begin{tabular}{r|c|c c}
\multicolumn{1}{r}{}
 &  \multicolumn{1}{c}{\scriptsize All} & \multicolumn{1}{c}{\scriptsize Unique}  & \multicolumn{1}{c}{\scriptsize }\\
\cline{2-3}
\scriptsize  & \tiny 1,955 & \multicolumn{1}{c|}{\cellcolor[gray]{0.8} \tiny 355=\textbf{18.16\%} } & \\
\cline{2-3}
\multicolumn{1}{r}{}
\underline{\textbf{\emph{Syria}}} &  \multicolumn{1}{c}{\tiny Archived} & \multicolumn{1}{c}{\tiny Not Archived} & \multicolumn{1}{c}{\scriptsize }\\
\cline{2-3}
\tiny Available & \tiny 19=5.35\% & \multicolumn{1}{c|}{\tiny 311=87.61\% }  & \tiny\\
\cline{2-4}
\tiny Missing & \tiny 0=0\% & \multicolumn{1}{c|}{\tiny 25=7.04\% }  & \multicolumn{1}{c|}{\cellcolor[gray]{0.8} \tiny 25=\textbf{7.04\%} }  \\ 
\cline{2-4}
\scriptsize  &\cellcolor[gray]{0.8} \tiny 19=\textbf{5.35\%} & \tiny each/355 & \tiny \\
\cline{2-2}
\end{tabular}
\end{minipage}
      \vskip10pt
      \caption{Percentages of unique resources from all the extracted ones we obtained per event and the percentages of presence of those unique resources on live web and in archives. All resources = 21,625, Unique resources = 11,051}
      \label{tab:all}
\vskip-15pt
\end{table}
\subsection{Existence on the Live-Web}
After obtaining the unique URIs from the previous step we resolve all of them and classify them as Success or Failure. The \textit{Success} class includes all the resources that ultimately return a ``200 OK'' HTTP response. The \textit{Failure} class includes all the 
resources that return a ``4XX'' family response 
like: ``404 Not Found'', ``403 Forbidden'' and ``410 Gone'', the ``30X'' redirect family while having infinite loop redirects, and server errors with response ``50X''. To avoid 
transient errors we repeated the requests, on all datasets, several times for a week to resolve those errors.\\

We also test for ``Soft 404s'', which are pages that return ``200 OK'' response code but are not a representation of the resource, using a technique based on a heuristic for automatically discovering soft 404s from Bar-Yossef et al. \cite{Bar-yossef}. We also 
include no response from the server, as well as DNS timeouts, as failures. Note that failure means that this resource is \textbf{\textit{missing}} on the live web. Table \ref{tab:all} summarizes, for each dataset, the total percentages of the 
resources missing from the live web and the number of missing resources divided by the total number of unique resources.\\

\subsection{Existence in the Archives}
In the previous step we tested the existence of the unique list of URIs for each event on the live web. Next, we evaluate how many URIs have been archived in public web archives. To check those archives we utilize the Memento framework. 
If there is a memento for the URI, we download its memento timemap and analyze it. The timemap is a datestamp ordered list of all known archived versions (called ``mementos'') of a URI. Next, we parse this timemap and extract the number of mementos that point to versions of the 
resource in the public archives. We declare the resource to be archived if it has at least one memento. This step was also repeated several times to avoid the transient states of the 
archives before deeming a resource as unarchived. The results of this experiment along with the archive 
coverage percentage are presented in Table \ref{tab:all}.

\section{Existence as a Function of Time}
\begin{figure}[htb]
\vskip-23pt
\centering
\includegraphics[width=1\textwidth]{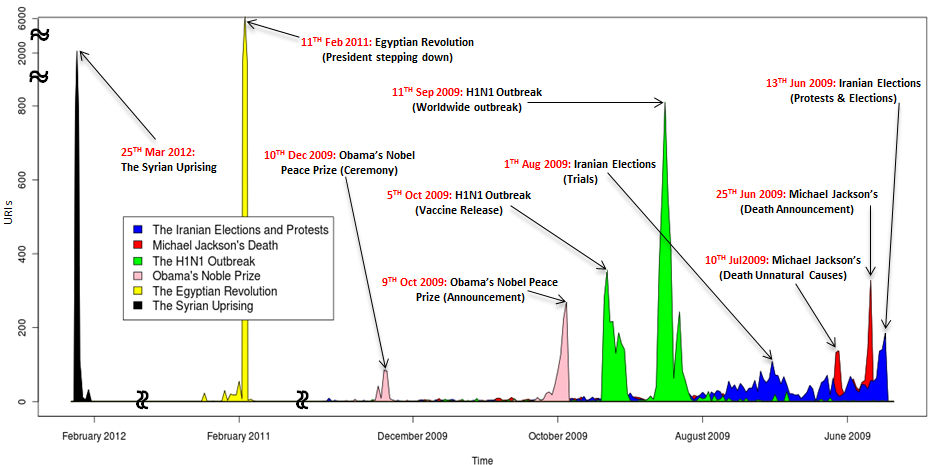}
\vskip-17pt
\caption{URIs shared per day corresponding to each event and showing the two peaks in the non-Syrian and non-Egyptian events}
\label{fig:allevents}
\vskip-15pt
\end{figure}

Inspecting the results from the previous steps suggests that the number of missing shared resources in social media corresponding to an event is directly proportional with its age. To determine dates for each of the events this we extracted all the 
creation dates from all the tweet-based datasets and sorted them. For each event, we plotted a graph illustrating the number of tweets per day related to that event as shown in figure \ref{fig:allevents}. Since the dataset is separated temporally into 
3 partitions, and in order to display all the events on one graph we reduced the size of the x-axis by removing the time periods not covered in our study.

Upon examining the graph we found an interesting phenomena in the non-Syrian and non-Egyptian events: each event has two peaks. Upon investigating history timelines we came to conclusion 
that those peaks reflect a second wave of social media interaction as a result of new incident within the same event after a period of time. For example, in the H1N1 dataset, the first peak illustrates the world-wide outbreak announcement while 
the second peak denotes the release of the vaccine. In the Iran dataset, the first peak shows the peak of the elections while the second peak pinpoints the Iranian trials. As for the MJ dataset the first peak corresponds to his death and the second peak 
describes the rumors that Michael Jackson died of unnatural causes and a possible homicide. For the Obama dataset, the first peak reveals the announcement of his winning the prize while the second peak presents the award-giving ceremony in 
Oslo. For the Egyptian evolution, the resources are all within a small time slot of 2 weeks around the date 11th of February. As for the Syrian event, since the collection was very recent there was no obvious peaks. Those peaks we examined will become temporal centroids of the social content collections (the datasets). 
MJ (June 25th \& July 10th 2009), Iran (June 13th \& 1st August 2009), H1N1 (September 11th \& 5th October 2009), and Obama (October 9th \& December 10th 2009). Egypt was (February 11th 2011) and the Syria dataset also had one centroid on March 27th 2012. 
We split each event according to the two centroids in each event accordingly. Figure \ref{fig:allevents} shows those peaks and Table \ref{tab:split} shows the missing content and the archived content percentages 
corresponding to each centroid.
\begin{center}
  \begin{table}
\vskip-23pt
      \begin{tabular}[c]{ | p{1.8cm}|| p{0.8cm} | p{0.8cm} || p{0.8cm} | p{0.8cm} || p{0.8cm} | p{0.8cm} || p{0.8cm} | p{0.8cm} || c || c |}
      \hline  
      &\multicolumn{2}{|c||}{\textbf{  MJ  }} & \multicolumn{2}{|c||}{\textbf{   Iran   }} & \multicolumn{2}{|c||}{\textbf{  H1N1  }} &  \multicolumn{2}{|c||}{\textbf{ 
Obama  }} & \textbf{ Egypt  } & \textbf{ Syria }\\ \hline \hline
      \scriptsize \% Missing & \tiny 36.24\% & \tiny  31.62\% & \tiny  26.98\%  & \tiny 24.47\% & \tiny 23.49\% & \tiny  25.64\% & \tiny  24.59\%  & \tiny 26.15\% & \tiny  10.48\% & \tiny 7.04\% \\ \hline
      \scriptsize \% Archived & \tiny 39.45\% & \tiny  30.78\% & \tiny  43.08\%  & \tiny 36.26\% & \tiny 41.65\% & \tiny  43.87\% & \tiny  47.87\%  & \tiny 46.15\% & \tiny  20.18\% & \tiny 5.35\% \\ \hline
      \end{tabular}\\ \normalsize 
      \caption{The Split Dataset}
      \label{tab:split}
\vskip-50pt
  \end{table}
\end{center}
\begin{figure}[htb]
\vskip-15pt
\centering
\includegraphics[width=1\textwidth]{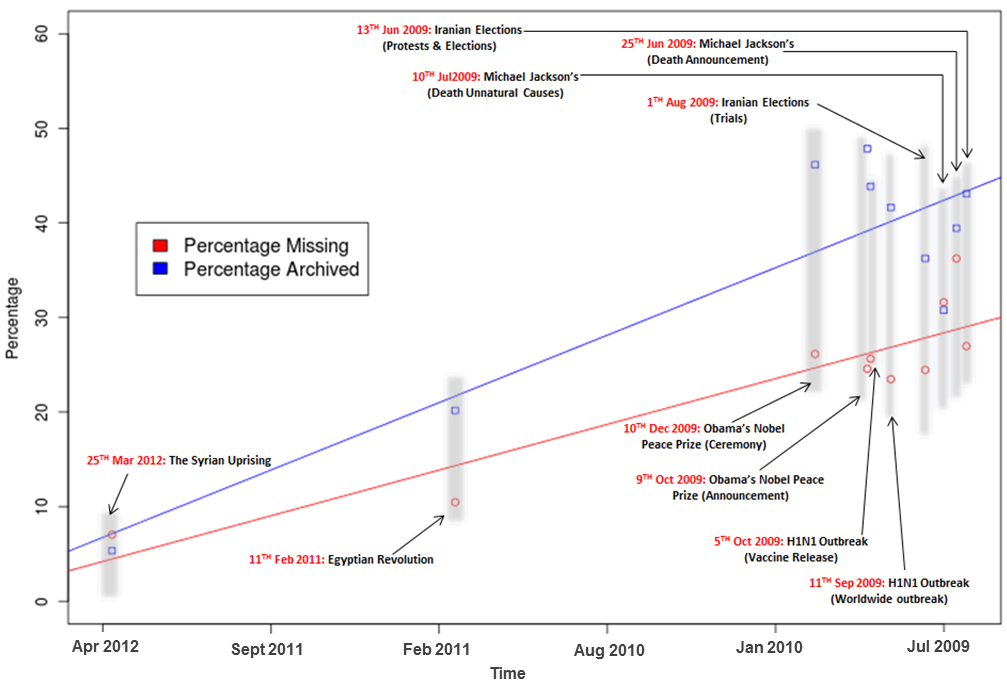}
\vskip-15pt
\caption{Percentage of content missing and archived for the events as a function of time.}
\vskip-20pt
\label{fig:thebigfigure}
\end{figure}
Figure \ref{fig:thebigfigure} shows the missing and archived values from Table \ref{tab:split} as a function of time since shared. Equation \ref{missing} shows the 
modeled estimate for the percentage of shared resources lost, where \textit{Age} is in days. While there is a less linear relationship between time and being archived, equation \ref{archived} shows the modeled estimate for the percentage of shared resources archived in a public archive.
\begin{equation}\label{missing}
Content\hspace{0.8 mm}Lost\hspace{0.8 mm}Percentage = 0.02 (Age\hspace{0.8 mm}in\hspace{0.8 mm}days) + 4.20
\end{equation}
\begin{equation}\label{archived}
Content\hspace{0.8 mm}Archived\hspace{0.8 mm}Percentage= 0.04 (Age\hspace{0.8 mm}in\hspace{0.8 mm}days) + 6.74
\end{equation}
Given these observations and our curve fitting we estimate that after a year from publishing about 11\% of content shared in social media will be gone. After this point, we are losing roughly 0.02\% of this content per day.

\section{Conclusions and Future work}
We can conclude that there is a nearly linear relationship between time of sharing in the social media and the percentage lost. Although not as linear, there is a similar relationship between the time of sharing and the expected 
percentage of coverage in the archives. To reach this conclusion, we extracted collections of tweets and other social media content that was posted and shared in relation to six different events that occurred in the time period from June 2009 to March 2012. Next we 
extracted the embedded resources within this social media content and tested their existence on the live web and in the archives. After analyzing the percentages lost and archived in relation to time and plotting them we used a linear regression model to fit 
those points. Finally we presented two linear models that can estimate the existence of a resource, that was posted or shared at one point of time in the social media, on the live web and in the archives as a function of age in the social media.\\

In the next stage of our research we need to expand the datasets and import other similar datasets especially in the uncovered temporal areas (e.g., the year of 2010 and before 2009). Examining more datasets across extended points in time could enable 
us to better model these two functions of time. Also several other factors beside time would be analyzed to understand their effect on persistence on the live web and archiving coverage like: publishing venue, rate of sharing, popularity of authors and the nature of the related event.
\vskip-25pt
\section{Acknowledgments}
This work was supported in part by the Library of Congress and NSF IIS-1009392.


\vskip-25pt
\begin{thebibliography}{4}

\bibitem{Ainsworth:2011:MWA:1998076.1998100}
Ainsworth, Scott G. and Alsum, Ahmed and SalahEldeen, Hany and Weigle, Michele C. and Nelson, Michael L.:
How Much of the Web Is Archived?
In \textit{Proceedings of the 11th annual international ACM/IEEE joint conference on Digital libraries, JCDL '11}, pages 133-136, (2011).

\bibitem{Bar-yossef}
Bar-Yossef, Ziv and Broder, Andrei Z. and Kumar, Ravi and Tomkins, Andrew.:
Sic Transit Gloria Telae: Towards an Understanding of the Web's Decay.
In \textit{Proceedings of the 13th international conference on World Wide Web, WWW '04}, pages 328-337, (2004).

\bibitem{Benevenut}
F. Benevenut, T. Rodrigues, M. Cha, and V. Almeida.:
Characterizing User Behavior in Online Social Networks.
In \textit{In Proc. of ACM SIGCOMM Internet Measurement Conference, SIGCOMM '09}, pages 49-62, (2009).


\bibitem{Chei}
Lee, Chei and Ma, Long and Goh, Dion.:
Why Do People Share News in Social Media?
\textit{Active Media Technology}, Springer Berlin / Heidelberg, pages 129-140, Volume:6890, (2011).

\bibitem{facebook} Facebook official fact sheet, \url{http://newsroom.fb.com/content/default.aspx?NewsAreaId=22}

\bibitem{Kwak} 
Kwak, Haewoon and Lee, Changhyun and Park, Hosung and Moon, Sue.:
What is Twitter, a Social Network or a News Media?
In \textit{Proceedings of the 19th international conference on World wide web, WWW '10}, pages 591-600, (2010).

\bibitem{heritrix} 
Gordon Mohr, Michele Kimpton, Micheal Stack and Igor Ranitovic.:
Introduction to Heritrix, an Archival Quality Web Crawler.
In \textit{4th International Web Archiving Workshop, IWAW '04},(2004).


\bibitem{1255182}
Frank McCown and Norou Diawara and Michael L. Nelson.:
Factors Affecting Website Reconstruction from the Web Infrastructure.
In \textit{Proceedings of the 7th ACM/IEEE-CS Joint Conference on Digital Libraries, JCDL '07}, pages 39-48, (2007).

\bibitem{nelson1} 
Michael L. Nelson, B. Danette Allen.:
Object Persistence and Availability in Digital Libraries.
\textit{D-Lib Magazine}, Volume 8, Number 1, January (2002)

\bibitem{human} 
M. E. J. Newman and J. Park.:
Why social networks are different from other types of networks. Phys. Rev. E, 68(3):036122, September, (2003).

\bibitem{tweetsfromtahrir} Alex Nunns and Nadia Idle.: Tweets From Tahrir. ISBN-10: 1935928457.

\bibitem{lexical}
T. A. Phelps and R. Wilensky.:
Robust Hyperlinks Cost Just Five Words Each. 
\textit{Technical Report}, UCB/CSD-00-1091, 
EECS Department, University of California, Berkeley, (2000).


\bibitem{losingrevolution} Hany M. SalahEldeen, Michael L. Nelson.: Losing My Revolution: A year after the Egyptian Revolution, 10\% of the social media documentation is gone. \url{http://ws-dl.blogspot.com/2012/02/2012-02-11-losing-my-revolution-year.html}

\bibitem{Sanderson} 
Robert Sanderson, Mark Phillips and Herbert Van de Sompel.:
Analyzing the Persistence of Referenced Web Resources with Memento.
\textit{CoRR}, arXiv:1105.3459, (2011)

\bibitem{snap} Stanford SNAP Project Dataset, \url{http://snap.stanford.edu/}

\bibitem{Twitternums} Twitter numbers, \url{http://blog.Twitter.com/2011/03/numbers.html}

\bibitem{memento}
H. Van de Sompel, M. L. Nelson, R. Sanderson, L. L. Balakireva, S. Ainsworth, H. Shankar.: Memento: Time Travel for the Web, Technical Report, arXiv:0911.1112, November, (2009).

\bibitem{lexical2} 
Wan, X., Yang, J.: Wordrank-based Lexical Signatures for Finding Lost or Related Web Pages.
In \textit{Proceedings of the 8th Asia-Pacific Web conference on Frontiers of WWW Research and Development, APWeb'06}, pages 843-849, (2006).

\bibitem{Wilson}
C. Wilson, B. Boe, A. Sala, K. P. Puttaswamy, and B. Y. Zhao.: 
User Interactions in Social Networks and their Implications. 
In \textit{Proceedings of the 4th ACM European conference on Computer systems, EuroSys '09}, pages 205-218, (2009).

\bibitem{Wu} 
Wu, Shaomei and Hofman, Jake M. and Mason, Winter A. and Watts, Duncan J.:
Who Says What to Whom on Twitter.
In \textit{Proceedings of the 20th international conference on World wide web, WWW '11}, pages 705-714, (2011).

\bibitem{Twitter}
Jaewon Yang and Jure Leskovec.:
Patterns of Temporal Variation in Online Media.
In \textit{ACM International Conference on Web Search and Data Minig, WSDM '11}, pages 177-186, (2011).

\bibitem{Yang} 
J. Yang and S. Counts.:
Predicting the Speed, Scale, and Range of Information Diffusion in Twitter.
In \textit{4th International AAAI Conference on Weblogs and Social Media, ICWSM '10}, May, (2010).

\bibitem{Zhao} 
D. Zhao and M. B. Rosson.:
How and Why People Twitter: The Role that Micro-blogging Plays in Informal Communication at Work.
In \textit{Proceedings of the ACM 2009 international conference on Supporting group work. GROUP '09}, pages 243-252, (2009).









\end{thebibliography}
\end{document}